\documentclass[5p,fleq30n,twocolumn,dvipfdmx]{elsarticle}
\usepackage{graphicx}
\usepackage{amsmath}
\usepackage{amssymb}
\usepackage{bm}
\usepackage{color}

\def\l{\langle}
\def\r{\rangle}

\newcommand{\IpyroTc}{4.21394(2)}
\newcommand{\Inu}{0.629(2)}
\newcommand{\IscTc}{4.511524(20)}
\newcommand{\Ibetanu}{0.520(5)}
\newcommand{\XYpyroTc}{2.02850(2)}
\newcommand{\XYnu}{0.672(2)}
\newcommand{\XYscTc}{2.20182(5)}
\newcommand{\XYbetanu}{0.515(5)}
\newcommand{\HpyroTc}{1.31695(2)}
\newcommand{\Hnu}{0.711(2)}
\newcommand{\HscTc}{1.4430(2)}
\newcommand{\Hbetanu}{0.515(5)}

\title{Large-scale calculation of ferromagnetic spin systems \\
on the pyrochlore lattice}
\author[fefu1]{Konstantin Soldatov\corref{cor1}}
\ead{soldatov_ks@students.dvfu.ru}
\author[fefu1,fefu2]{Konstantin Nefedev}
\ead{nefedev.kv@dvfu.ru}
\author[CIJ]{Yukihiro Komura}
\author[tmu]{Yutaka Okabe\corref{cor1}}
\ead{okabe@phys.se.tmu.ac.jp}
\address[fefu1]{School of Natural Sciences, Far Eastern Federal University,
Vladivostok, Russian Federation}
\address[fefu2]{Institute of Applied Mathematics, Far Eastern Branch,
Russian Academy of Science, Vladivostok, Russian Federation}
\address[CIJ]{CIJ-solutions, Chuo-ku, Tokyo, 103-0023, Japan}
\address[tmu]{Department of Physics, Tokyo Metropolitan University,
Hachioji, Tokyo 192-0397, Japan}

\cortext[cor1]{Corresponding author}

\begin{document}

\begin{abstract}
We perform the high-performance computation of the ferromagnetic Ising model
on the pyrochlore lattice.
We determine the critical temperature accurately based on
the finite-size scaling of the Binder ratio.
Comparing with the data on the simple cubic lattice, we argue
the universal finite-size scaling.
We also calculate the classical XY model and the classical Heisenberg model
on the pyrochlore lattice.
\end{abstract}


\maketitle

\section{Introduction}

Universality and scaling are two important concepts
in critical phenomena \cite{stanley71,Hu14}.  The critical phenomena
associated with the second-order phase transitions are
classified into a limited number of universality classes
defined not by detailed material parameters, but the fundamental
symmetries of a system, that is, the spatial dimension $D$,
the number of components of order parameter $n$, etc.

\begin{figure}[t]
\begin{center}
\includegraphics[width=7.2cm]{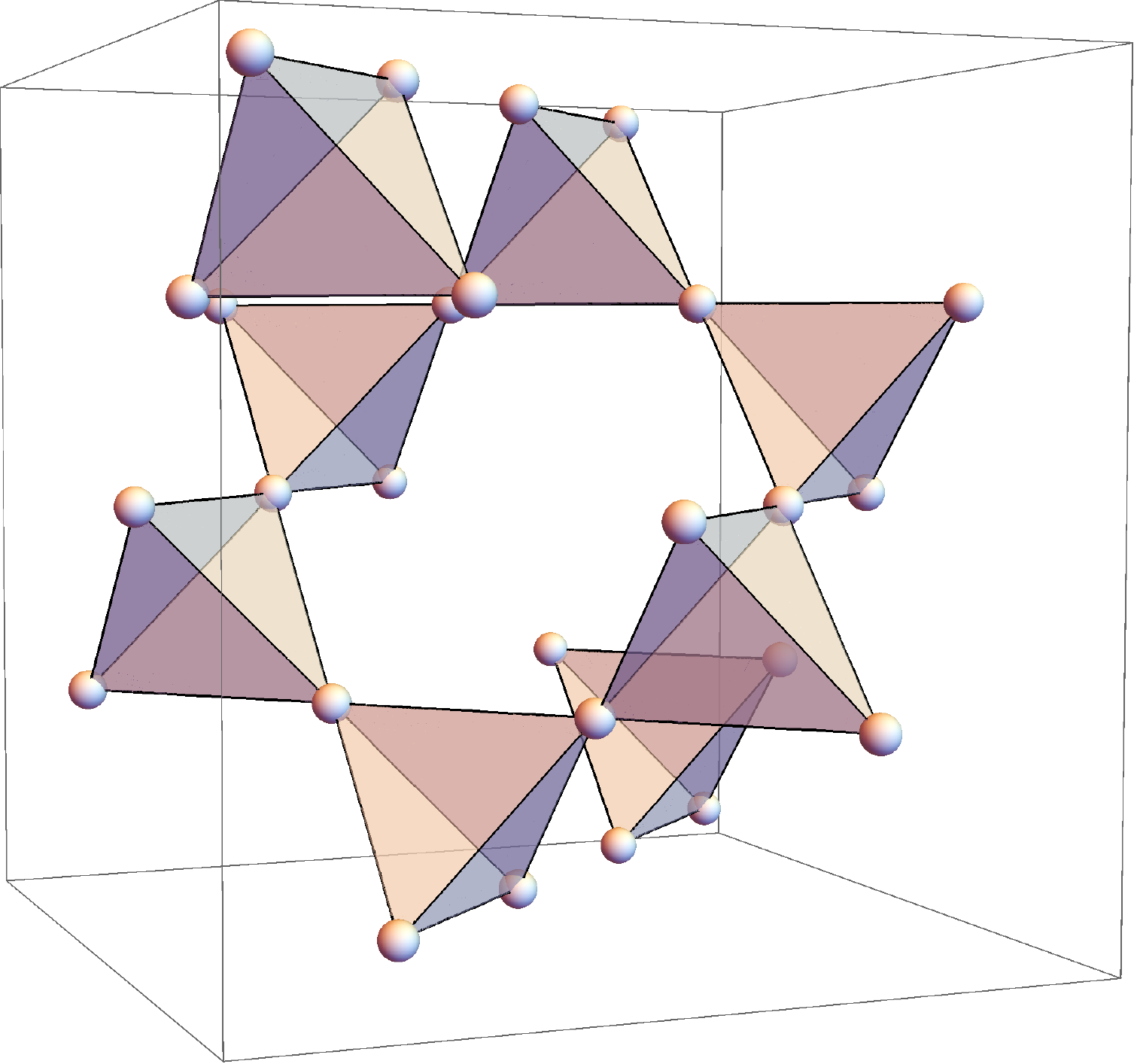}
\caption{\label{fig:fig1}
(Color online.)
The illustration of the pyrochlore lattice, which is a three-dimensional 
network of corner-sharing tetrahedra.
}
\end{center}
\end{figure}

In some problems, the lattice structure plays an important role.
Recently, the pyrochlore lattice has received
a lot of attention because of its relation to
the spin ice \cite{Harris,Ramirez,Bramwell}.
The pyrochlore lattice is a three-dimensional network of 
corner-sharing tetrahedra, and 
the illustration of the pyrochlore lattice is shown in Fig.~\ref{fig:fig1}.
Antiferromagnetic spin systems on the pyrochlore lattice
have frustration.
The dilution effects on frustration were also studied
for spin ice materials on the pyrochlore lattice \cite{Ke,Shevchenko}.
It is also interesting to study ferromagnetic spin systems
on the pyrochlore lattice in connection with the universality.

It is well known that Monte Carlo simulation is a standard method
to study statistical physics of many-body problems \cite{landau}.
The single spin flip Metropolis method \cite{metro53} is a robust algorithm
for a wide range of subjects, but often suffers from
the problem of slow dynamics; that is, it takes a long time
for equilibration, for example, at temperatures near the critical
temperature of the phase transition.  To conquer the problem
of slow dynamics, the cluster spin flip algorithms of Monte Carlo
simulation have been proposed.  The multi-cluster spin flip algorithm
due to Swendsen and Wang (SW) \cite{sw87} and the single-cluster
spin flip algorithm due to Wolff \cite {wolff89} are typical examples.

For high-performance computing, the use of graphic processing
unit (GPU) is a hot topic in computer science.
The parallelization of cluster spin flip algorithm
is not straightforward because the cluster labeling
part of the cluster spin flip algorithm basically requires a sequential
calculation, which is in contrast to the local calculation
for the single spin flip algorithm.
Komura and Okabe \cite{komura12} proposed the GPU computing
for the SW multi-cluster spin flip algorithm,
where the ideas of Hawick {\it et al.} \cite{Hawick_labeling}
and Kalentev {\it et al.} \cite{Kalentev}
were used in the cluster labeling part.
Recently, Komura \cite{komura15} proposed a refined version
of the SW multi-cluster spin flip algorithm with a single GPU.
Sample programs of the methods of Refs. \cite{komura12} and
\cite {komura15} were published \cite{komura_program,komura15_program}.
As applications, the large-scale Monte Carlo study of the two-dimensional
XY model \cite{komura_xy}, and the phase transitions of the Ising model
on the Penrose lattice (quasicrystal) \cite{komura_penrose} were studied.

In this paper, we perform the high-performance computation of
the ferromagnetic Ising model on the pyrochlore lattice.
We use the GPU algorithm by Komura \cite{komura15}
for the SW method.  We determine the critical temperature
accurately based on the finite-size scaling (FSS) \cite{fisher70}
of the Binder ratio \cite{Binder}.
Comparing with the data on the simple cubic lattice,
we argue the universal FSS \cite{Hu95,Okabe96}.
We also calculate the classical XY model and the
classical Heisenberg model on the pyrochlore lattice.

The remaining part of the paper is organized
as follows: The model and the method are described in Sec.~2.
The results are 
presented and 
discussed in Sec.~3, while Sec.~4 is devoted
to the concluding remark.

\section{Model and Simulation Method}

We deal with the classical spin models on the pyrochlore lattice.
For the simulation, we use the 16-site cubic unit cell
of the pyrochlore lattice \cite{Shimaoka}, and the systems
with $L \times L \times L$ unit cells with periodic boundary conditions
are treated.  We made simulations for the system sizes up to
$L$ =96; the numbers of sites are $N$ (=$16 L^3$) = 14155776.

The Hamiltonian of the classical spin models is given by
\begin{equation}
 \mathcal{H} = -J \sum_{\l i,j \r} \bm{s}_i \cdot \bm{s}_j,
\end{equation}
where $J$ is the coupling and $\bm{s}_i$ is
an $n$-dimensional unit vector on the lattice site $i$;
$n$=1, 2, and 3 correspond to the Ising model,
the classical XY model, and the classical Heisenberg model,
respectively.
The summation is taken over the nearest-neighbor pairs
$\l i,j \r$.  We note that the coordination number of
the pyrochlore lattice is six, which is the same as the simple
cubic lattice.

We use the SW multi-cluster spin flip algorithm with a single GPU
in the program of Ref.~\cite{komura15,komura15_program}.
The embedded cluster idea of Wolff \cite{wolff89} is used
for simulating the spin systems with continuous symmetry
(XY model and Heisenberg model).
Once we have the table which gives the nearest-neighbor sites
for each site, we can use the CUDA program of
Ref.~\cite{komura15,komura15_program}. 
The performance of the parallel computation with using the 
CUDA program was discussed in Ref.~\cite{komura15,komura15_program}.
The system sizes we treat are $L$ = 32 ($N$ = 524288), $L$ = 48
($N$ = 1769472), $L$ = 64 ($N$ = 4194304), and $L$ = 96 ($N$ = 14155776).
We discarded the first 10,000 Monte Carlo Steps (MCSs) 
to avoid the effects of initial configurations, 
and the next 200,000 MCSs were used for measurement.
We made five independent runs for each size; the average was
taken over five runs, and the statistical errors were estimated.

\section{Results}

\subsection{Ising model}

\begin{figure}
\begin{center}
\includegraphics[width=7.2cm]{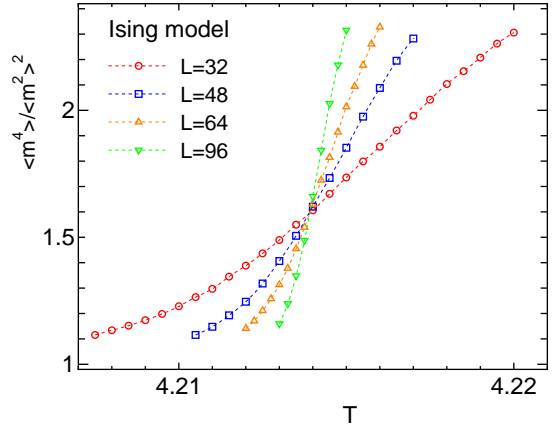}
\caption{\label{fig:fig2}
(Color online.) The moment ratio of the Ising model on the pyrochlore lattice.
The system sizes are $L=32 \ (N=524288)$, $L=48 \ (N=1769472)$,
$L=64 \ (N=4194304)$, and $L=96 \ (N=14155776)$.
}
\end{center}
\end{figure}

We start with the Ising model on the pyrochlore lattice.
This model undergoes a second-order phase transition.
To study the critical phenomena of second-order phase transition, 
it is convenient to calculate the moment ratio of 
the magnetization $M$ \cite{Binder}.
In Fig.~\ref{fig:fig2}, we plot the temperature dependence
of the moment ratio of the magnetization $M$;
\begin{equation}
   U(T) = \frac{\l M(T)^4 \r}{\l M(T)^2 \r^2}
        = \frac{\l m(T)^4 \r}{\l m(T)^2 \r^2}
\end{equation}
with $m=M/N$, which is essentially the Binder ratio \cite{Binder}
except for the normalization.
The value of $U(T)$ becomes 1 for $T \to 0$,
whereas it becomes 3 for $T \to \infty$.
In the high temperature limit the fluctuations become Gaussian, 
and a simple calculation yields $U(T) \to (n+2)/n$. 
The system sizes are $L=32 \ (N=524288)$, $L=48 \ (N=1769472)$,
$L=64 \ (N=4194304)$, and $L=96 \ (N=14155776)$.
The temperature is measured in units of $J$; in other words,
we take $J=1$. The error bars are within the size of marks.
We only show the data near the second-order phase transition.
We see from Fig.~\ref{fig:fig2} that the data with different $L$
cross around $T=4.214$, which yields the critical temperature $T_c$.

\begin{figure}
\begin{center}
\includegraphics[width=7.2cm]{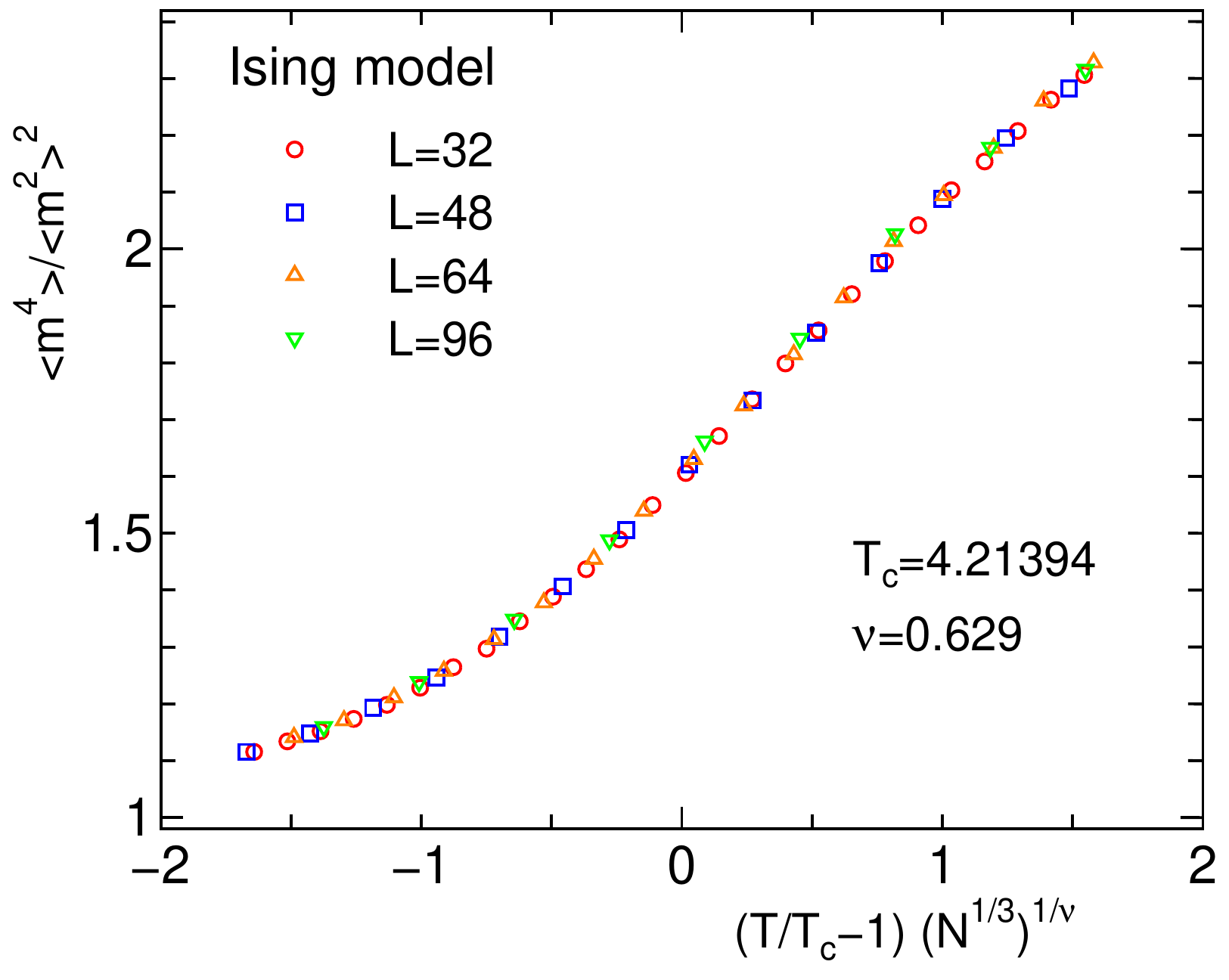}
\caption{\label{fig:fig3}
(Color online.) The FSS plot of the moment ratio of the Ising model
on the pyrochlore lattice. The system size $N$ is $16 L^3$.
The choices of $T_c$ and $\nu$ are given in the figure.
}
\end{center}
\end{figure}

To get precise estimate of $T_c$, let us consider the FSS \cite{fisher70}
of the moment ratio $U(T)$.
The FSS of $U(T)$ is expected to take a form
\begin{equation}
   U(t) = f_1 \Big(t \, (N^{1/3})^{1/\nu} \Big),
\label{FSS_1}
\end{equation}
where $t=(T-T_c)/T_c$, and $\nu$ is the critical exponent
for the correlation length.
We plot $U(T)=\l m^4 \r/\l m^2 \r^2$ as a function of $t(N^{1/3})^{1/\nu}$
in Fig.~\ref{fig:fig3};
all the data with different sizes are collapsed on a single curve
within statistical errors.
As for the linear size, we use $N^{1/3} = 16^{1/3} L$
for a later convenience of the discussion
of universal FSS.
Here, the best choices of $T_c$ and $\nu$ are \IpyroTc\ and \Inu,
respectively.
The estimated critical exponent $\nu$ is a universal one
of the three-dimensional (3D) Ising exponent~\cite{blote}.
The estimated $T_c$, \IpyroTc, is about 93.4\% of $T_c$
of the simple cubic lattice,
\IscTc\ (Ref.~\cite{blote}), although the coordination number
of both lattices is the same, that is, six.

\begin{figure}
\begin{center}
\includegraphics[width=7.2cm]{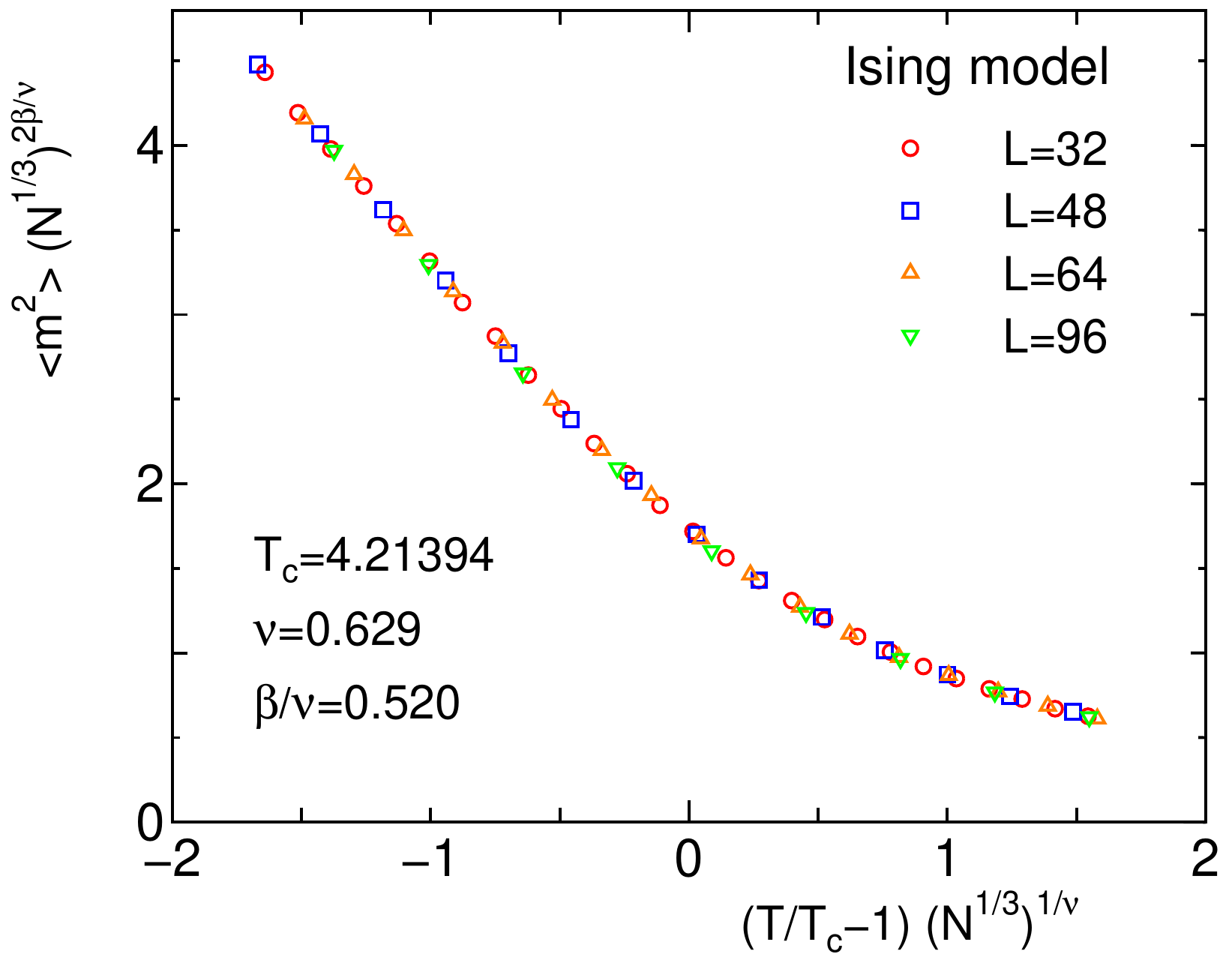}
\caption{\label{fig:fig4}
(Color online.) The FSS plot of the squared magnetization of the Ising model
on the pyrochlore lattice. The system size $N$ is $16 L^3$.
The choices of $T_c$, $\nu$, and $\beta/\nu$ are given in the figure.
}
\end{center}
\end{figure}

We also consider the FSS of the magnetization.
The FSS form of the squared magnetization is
\begin{equation}
  \l m^2 \r = (N^{1/3})^{-2\beta/\nu} f_2 \Big(t \, (N^{1/3})^{1/\nu} \Big),
\label{FSS_2}
\end{equation}
where $\beta$ is the critical exponent for the magnetization.
In Fig.~\ref{fig:fig4}, we give the FSS plot of
the second moment of the magnetization of the Ising model
on the pyrochlore lattice.
We use the same $T_c$ and $\nu$ as the FSS plot of
the moment ratio, and the best choice of $\beta/\nu$ is \Ibetanu.
We see that the FSS works very well.

\subsection{XY model}

\begin{figure}
\begin{center}
\includegraphics[width=7.2cm]{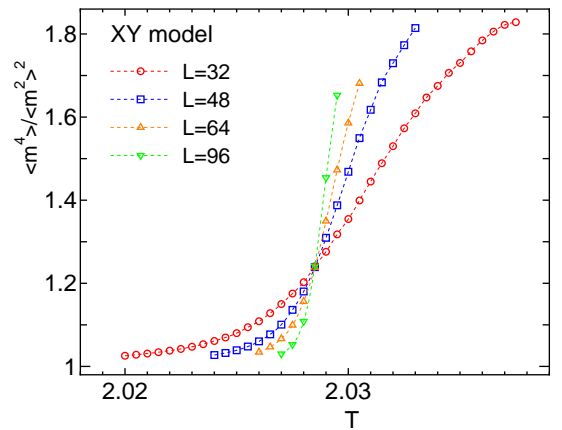}
\caption{\label{fig:fig5}
(Color online.) The moment ratio of the XY model on the pyrochlore lattice.
The system sizes are $L=32 \ (N=524288)$, $L=48 \ (N=1769472)$,
$L=64 \ (N=4194304)$, and $L=96 \ (N=14155776)$.
}
\end{center}
\end{figure}

We turn to the classical XY model on the pyrochlore lattice.
The temperature dependence of the moment ratio of
the classical XY model on the pyrochlore lattice
is plotted in Fig.~\ref{fig:fig5}.
In the case of the XY model ($n=2$), $U(T)$ becomes 2
for $T \to \infty$.
The system sizes are the same as the case of the Ising model.
The error bars are within the size of marks.
We see from the figure that the data with different $L$
cross around $T=2.029$.

\begin{figure}
\begin{center}
\includegraphics[width=7.2cm]{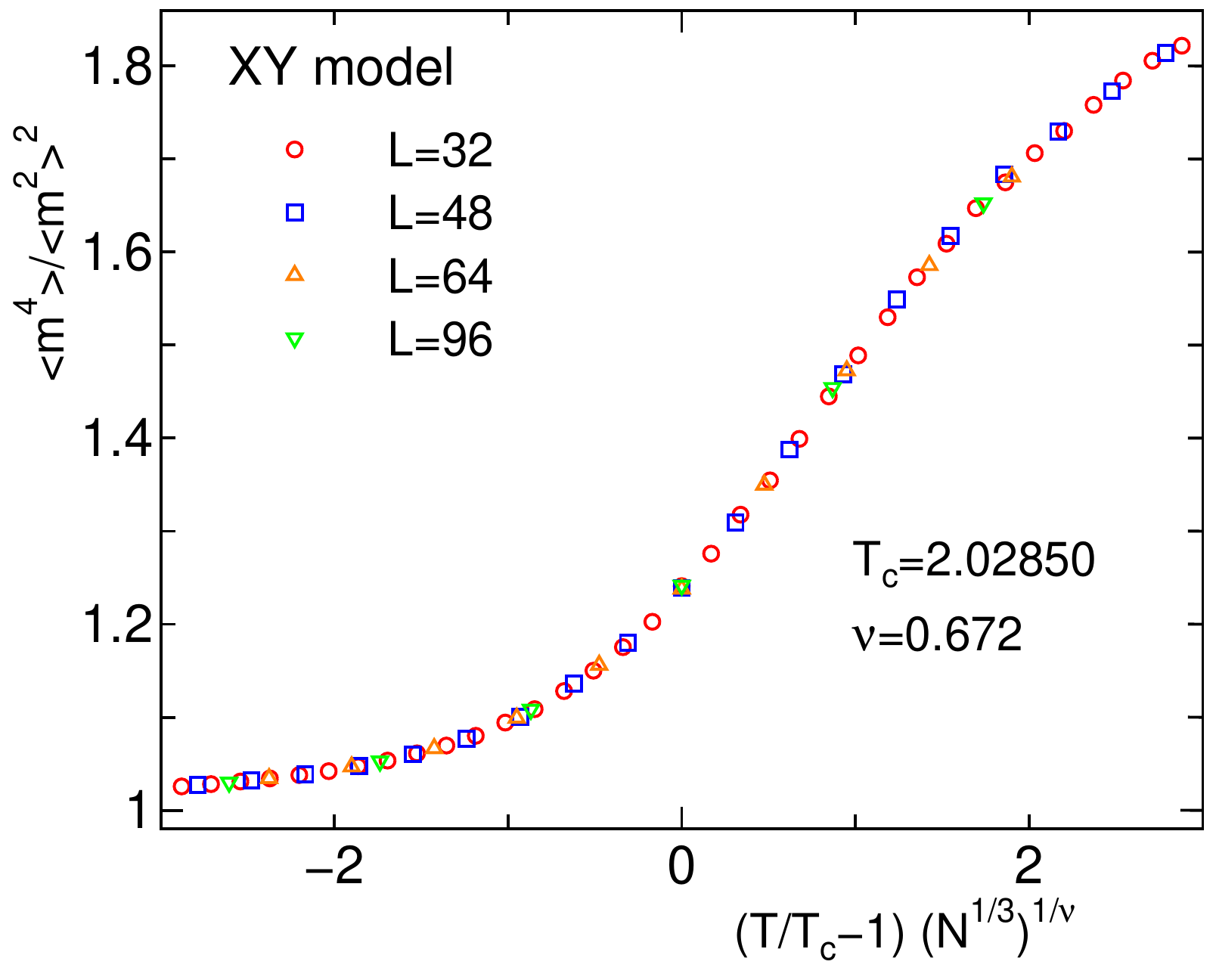}
\caption{\label{fig:fig6}
(Color online.) The FSS plot of the moment ratio of the XY model
on the pyrochlore lattice. The system size $N$ is $16 L^3$.
The choices of $T_c$ and $\nu$ are given in the figure.
}
\end{center}
\end{figure}

In Fig.~\ref{fig:fig6}, we give the FSS plot of
the moment ratio of the classical XY model
on the pyrochlore lattice.
The data collapsing of different sizes is very good again.
The best choices of $T_c$ and $\nu$ are \XYpyroTc\ and \XYnu,
respectively.
The estimated critical exponent $\nu$ is a universal one
of the 3D XY exponent~\cite{Campostrini2001}.
The estimated $T_c$, \XYpyroTc, is about 92.1\% of $T_c$
of the simple cubic lattice, \XYscTc\ (Ref.~\cite{Gottlob}).

\begin{figure}
\begin{center}
\includegraphics[width=7.2cm]{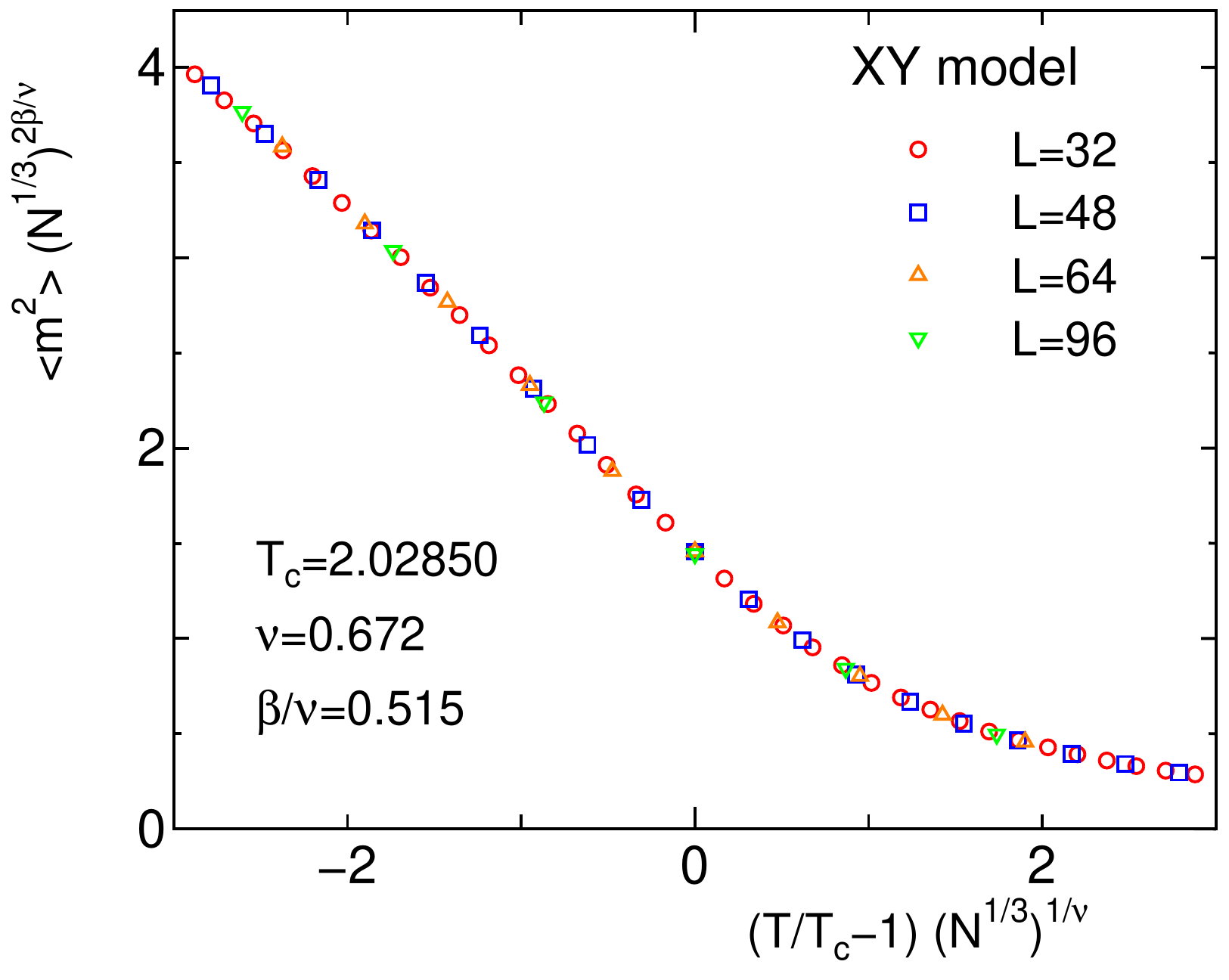}
\caption{\label{fig:fig7}
(Color online.) The FSS plot of the squared magnetization of the XY model
on the pyrochlore lattice. The system size $N$ is $16 L^3$.
The choices of $T_c$, $\nu$, and $\beta/\nu$ are given in the figure.
}
\end{center}
\end{figure}

The FSS plot of the second moment of the magnetization
of the classical XY model
on the pyrochlore lattice is shown in Fig.~\ref{fig:fig7}.
We use the same $T_c$ and $\nu$ as the FSS plot of
the moment ratio, and the best choice of $\beta/\nu$ is \XYbetanu.

\subsection{Heisenberg model}

\begin{figure}
\begin{center}
\includegraphics[width=7.2cm]{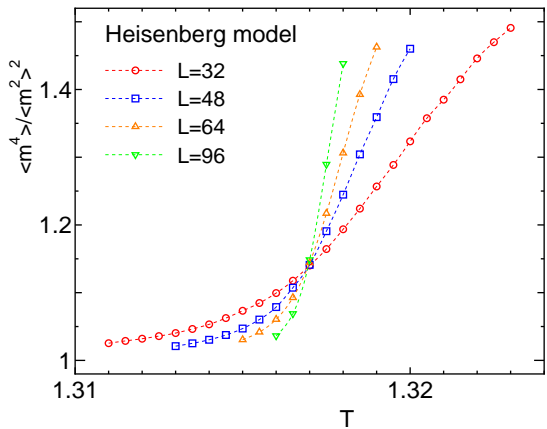}
\caption{\label{fig:fig8}
(Color online.) The moment ratio of the Heisenberg model on the pyrochlore
lattice. The system sizes are $L=32 \ (N=524288)$, $L=48 \ (N=1769472)$,
$L=64 \ (N=4194304)$, and $L=96 \ (N=14155776)$.
}
\end{center}
\end{figure}

Next, we consider the classical Heisenberg model on the pyrochlore lattice.
In Fig.~\ref{fig:fig8}, we plot the temperature dependence
of the moment ratio of the classical Heisenberg model
on the pyrochlore lattice.
In the case of Heisenberg model ($n=3$), $U(T)$ becomes 5/3
for $T \to \infty$.

\begin{figure}
\begin{center}
\includegraphics[width=7.2cm]{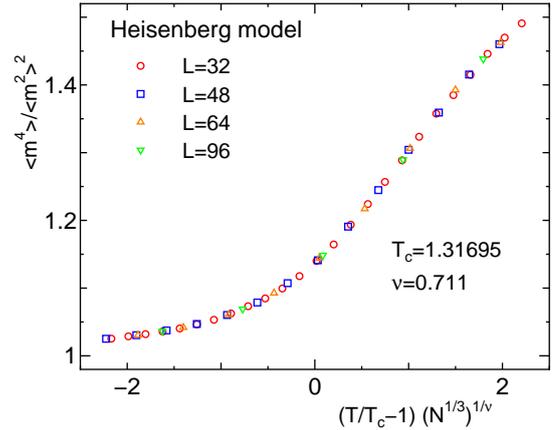}
\caption{\label{fig:fig9}
(Color online.) The FSS plot of the moment ratio of the Heisenberg model
on the pyrochlore lattice. The system size $N$ is $16 L^3$.
The choices of $T_c$ and $\nu$ are given in the figure.
}
\end{center}
\end{figure}

In Fig.~\ref{fig:fig9}, we show the FSS plot of
the moment ratio of the classical Heisenberg model
on the pyrochlore lattice.
The data collapsing of different sizes is very good again.
The best choices of $T_c$ and $\nu$ are \HpyroTc\ and \Hnu,
respectively.
The estimated critical exponent $\nu$ is a universal one
of the 3D Heisenberg exponent~\cite{Campostrini2002}.
The estimated $T_c$, \HpyroTc,
is about 91.3\% of $T_c$ of the simple cubic lattice,
\HscTc\ (Ref.~\cite{Holm}).

\begin{figure}
\begin{center}
\includegraphics[width=7.2cm]{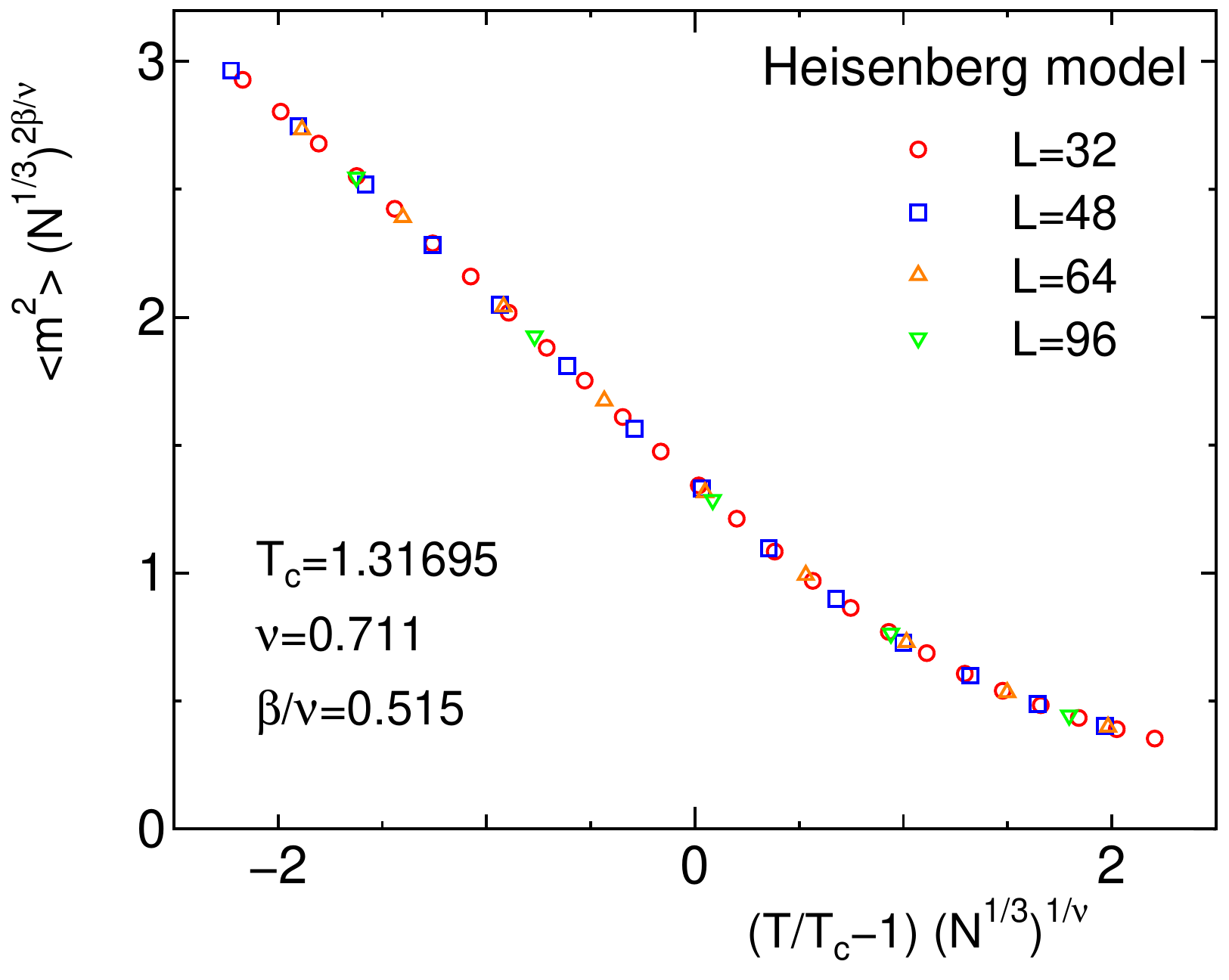}
\caption{\label{fig:fig10}
(Color online.) The FSS plot of the squared magnetization of the Heisenberg
model on the pyrochlore lattice. The system size $N$ is $16 L^3$.
The choices of $T_c$, $\nu$, and $\beta/\nu$ are given in the figure.
}
\end{center}
\end{figure}

The FSS plot of the second moment of the magnetization of
the classical Heisenberg model
on the pyrochlore lattice is given in Fig.~\ref{fig:fig10}.
We use the same $T_c$ and $\nu$ as the FSS plot of
the moment ratio, and the best choice of $\beta/\nu$ is \Hbetanu.

\subsection{Universal finite-size scaling}

The idea of universal FSS functions was reported for critical phenomena
in geometric percolation models \cite{Hu95}.
Hu, Lin, and Chen \cite{Hu95} applied a histogram Monte Carlo
simulation method to calculate the existence probability
$E_p$ and the percolation probability $P$ of site and bond percolation
on finite square, plane triangular, and honeycomb
lattices, whose aspect ratios approximately have the relative proportions
$1:\sqrt{3}/2:\sqrt{3}$.  They found that the six
percolation models have very nice universal FSS for $E_p$ and $P$
near the critical points of the percolation models with
using nonuniversal metric factors.
Using Monte Carlo simulation, Okabe and Kikuchi \cite{Okabe96} found
universal FSS for the Binder ratio \cite{Binder}
and magnetization distribution functions $p(m)$ of the Ising model
on planar lattices. 
The universal FSS was further discussed in connection with spin models 
\cite{Okabe99,Tomita99,Wu03}.

Here, we check the universal FSS for the classical spin systems
on the pyrochlore lattice and on the simple cubic lattice.
The FSS function for the Binder ratio, Eq.~(\ref{FSS_1}),
and that for the magnetization, Eq.~(\ref{FSS_2}), may
take the universal FSS forms such as
\begin{equation}
   U(T) = \frac{\l m(T)^4 \r}{\l m(T)^2 \r^2} = f_1(D_1t(N^{1/3})^{1/\nu}),
\end{equation}
\begin{equation}
   \l m^2 \r = D_2 (N^{1/3})^{-2\beta/\nu} f_2(D_1t(N^{1/3})^{1/\nu})
\end{equation}
with nonuniversal metric factors $D_1$ and $D_2$.

\begin{figure}
\begin{center}
\includegraphics[width=7.2cm]{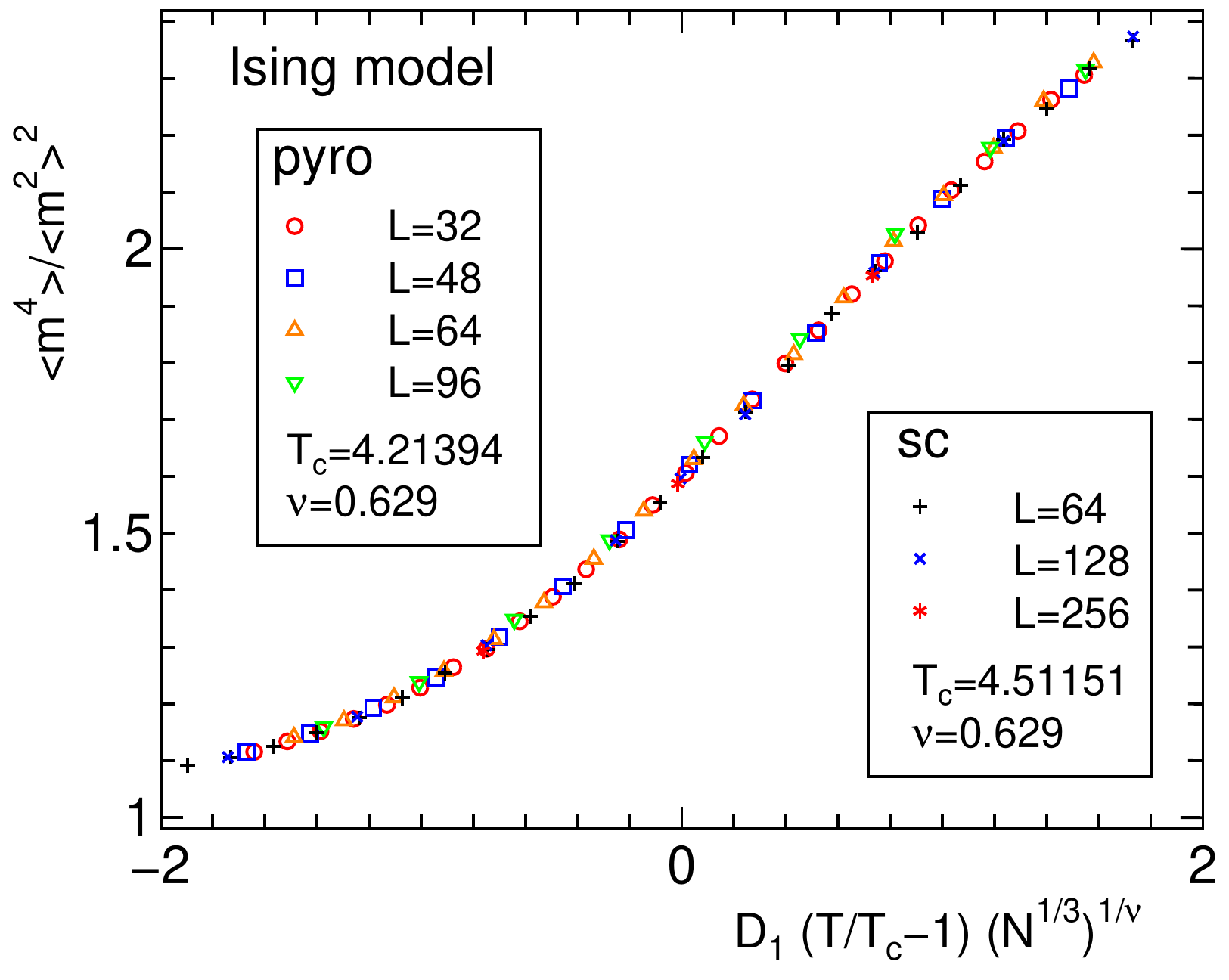}
\caption{\label{fig:fig11}
(Color online.) The universal FSS plot of the moment ratio of the Ising model.
The system size $N$ is $16 L^3$ for the pyrochlore lattice, whereas it is
$L^3$ for the simple cubic lattice.
The choices of $T_c$ and $\nu$ are given in the figure.
The nonuniversal metric factor $D_1$ is chosen as 1.0 for the simple cubic
lattice.  The estimated $D_1$ for the pyrochlore lattice is 1.0.
}
\end{center}
\end{figure}

\begin{figure}
\begin{center}
\includegraphics[width=7.2cm]{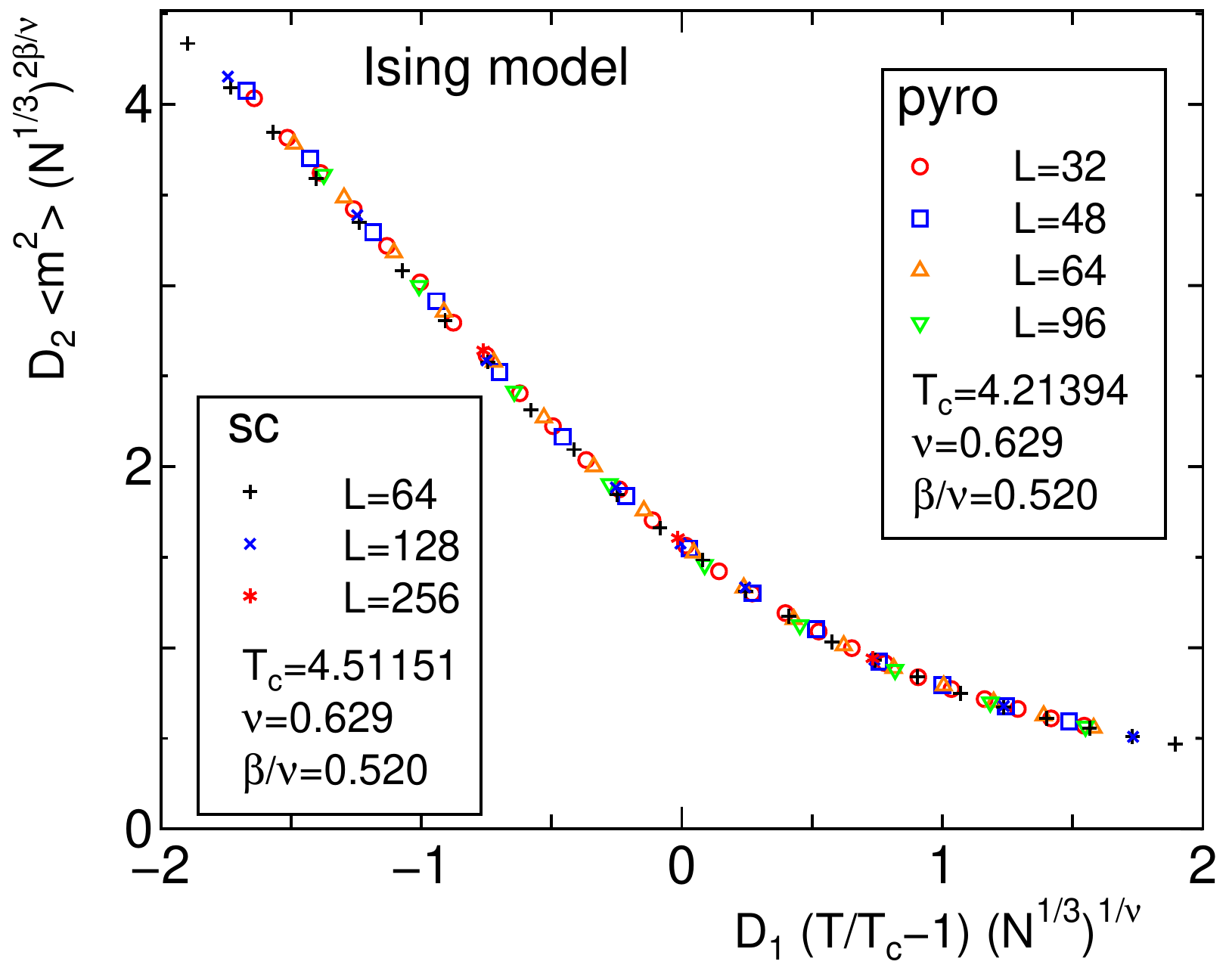}
\caption{\label{fig:fig12}
(Color online.) The universal FSS plot of the squared magnetization
of the Ising model.
The system size $N$ is $16 L^3$ for the pyrochlore lattice, whereas it is
$L^3$ for the simple cubic lattice.
The choices of $T_c$, $\nu$, and $\beta/\nu$ are given in the figure.
The nonuniversal metric factors $D_1$ and $D_2$ are chosen as 1.0
for the simple cubic lattice.
The estimated $D_1$ and $D_2$ for the pyrochlore lattice are 1.0 and
0.91, respectively.
}
\end{center}
\end{figure}

We show the universal FSS plot of the moment ratio of
the Ising model on the pyrochlore lattice and the simple cubic
lattice in Fig.~\ref{fig:fig11}.  The system size $N$ is $L^3$ with
the linear size $L$ for the simple cubic lattice.
The metric factor for the simple cubic lattice
is chosen as 1.0. The estimated metric factor $D_1$
for the pyrochlore lattice is 1.0.
We find that the universal FSS plot works very well.
The universal FSS plot of the squared magnetization of
the Ising model on the pyrochlore lattice and the simple cubic lattice
is given in Fig.~\ref{fig:fig12}.  The estimated another metric factor
$D_2$ is 0.91.

\begin{figure}
\begin{center}
\includegraphics[width=7.2cm]{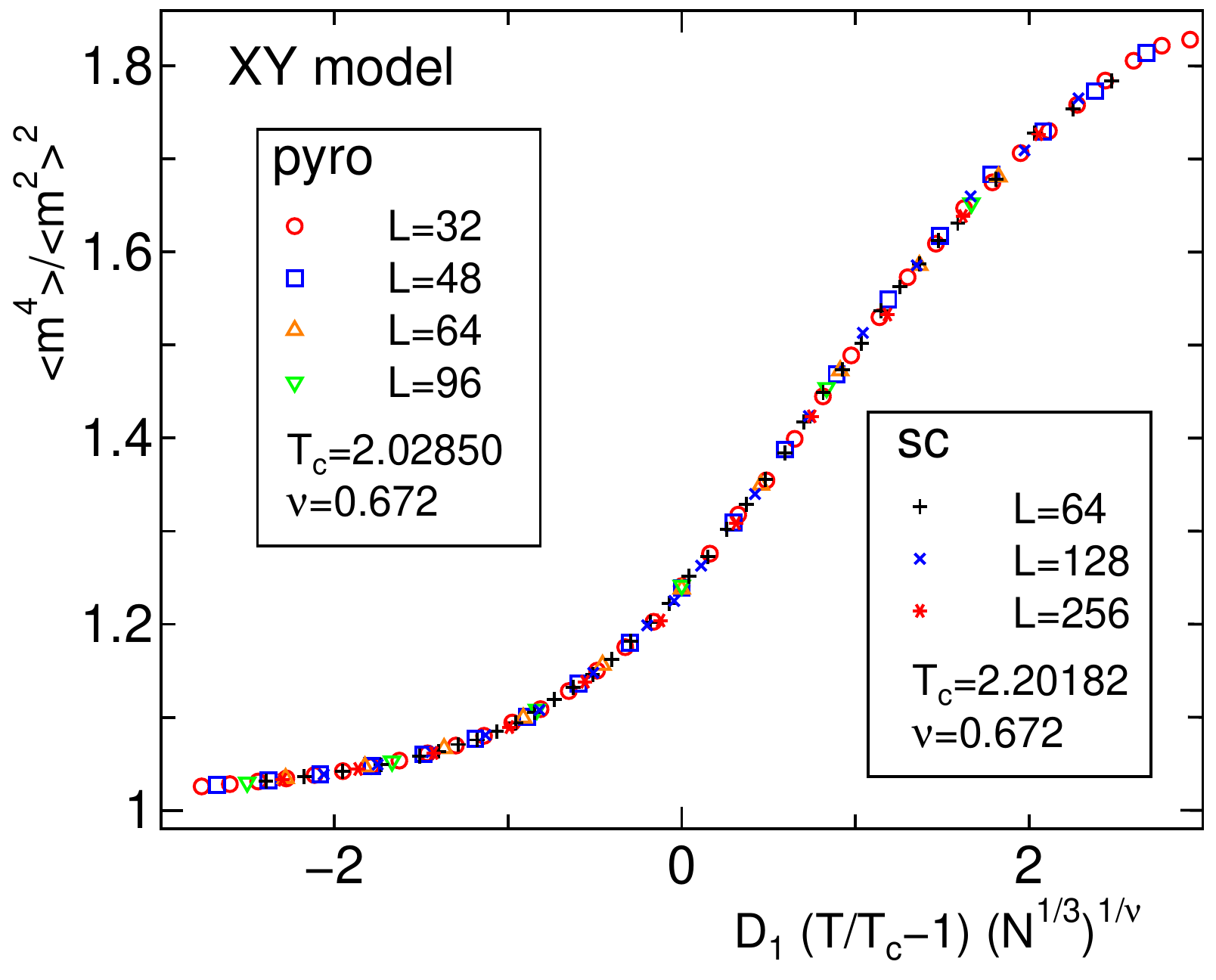}
\caption{\label{fig:fig13}
(Color online.) The universal FSS plot of the moment ratio of the XY model.
The system size $N$ is $16 L^3$ for the pyrochlore lattice, whereas it is
$L^3$ for the simple cubic lattice.
The choices of $T_c$ and $\nu$ are given in the figure.
The nonuniversal metric factor $D_1$ is chosen as 1.0 for the simple cubic
lattice.  The estimated $D_1$ for the pyrochlore lattice is 0.96.
}
\end{center}
\end{figure}

\begin{figure}
\begin{center}
\includegraphics[width=7.2cm]{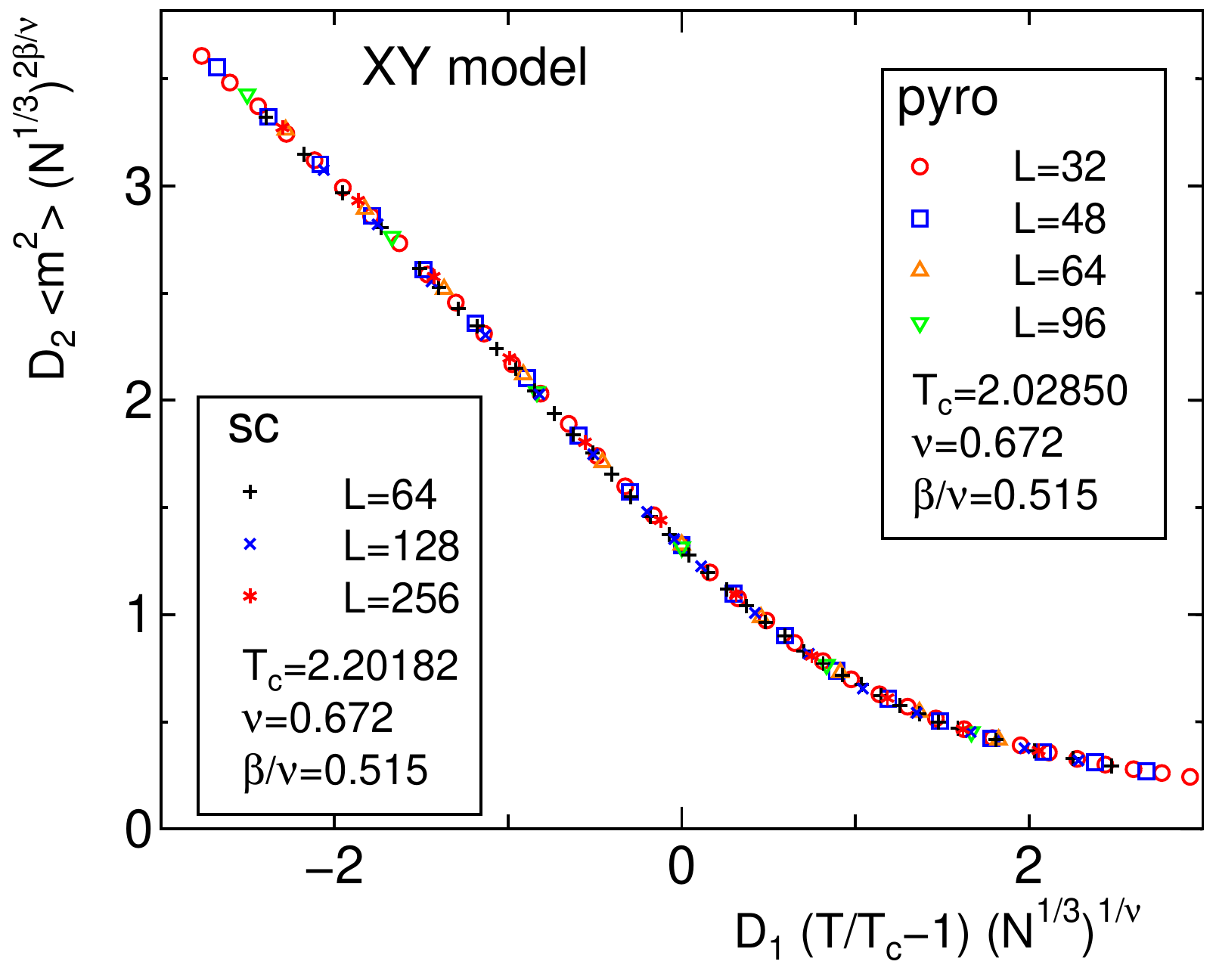}
\caption{\label{fig:fig14}
(Color online.) The universal FSS plot of the squared magnetization
of the XY model.
The system size $N$ is $16 L^3$ for the pyrochlore lattice, whereas it is
$L^3$ for the simple cubic lattice.
The choices of $T_c$, $\nu$, and $\beta/\nu$ are given in the figure.
The nonuniversal metric factors $D_1$ and $D_2$ are chosen as 1.0
for the simple cubic lattice.
The estimated $D_1$ and $D_2$ for the pyrochlore lattice are 0.96 and
0.91, respectively.
}
\end{center}
\end{figure}

The universal FSS plots of the moment ratio and of the squared
magnetization of the XY model are shown
in Figs.~\ref{fig:fig13} and \ref{fig:fig14}, respectively.
The nonuniversal metric factors
of the pyrochlore lattice are $D_1=0.96$ and $D_2=0.91$.

\begin{figure}
\begin{center}
\includegraphics[width=7.2cm]{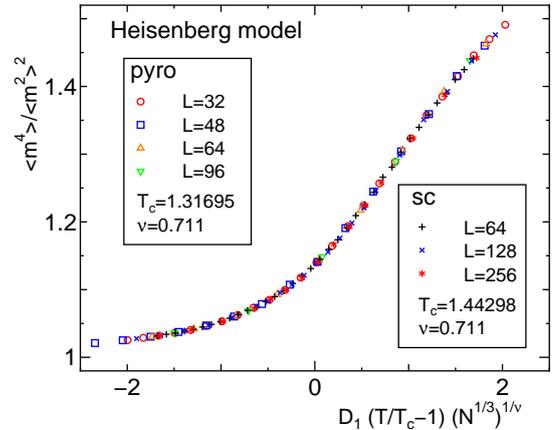}
\caption{\label{fig:fig15}
(Color online.) The universal FSS plot of the moment ratio of
the Heisenberg model.
The system size $N$ is $16 L^3$ for the pyrochlore lattice, whereas it is
$L^3$ for the simple cubic lattice.
The choices of $T_c$ and $\nu$ are given in the figure.
The nonuniversal metric factor $D_1$ is chosen as 1.0 for the simple cubic
lattice.  The estimated $D_1$ for the pyrochlore lattice is 0.92.
}
\end{center}
\end{figure}

\begin{figure}
\begin{center}
\includegraphics[width=7.2cm]{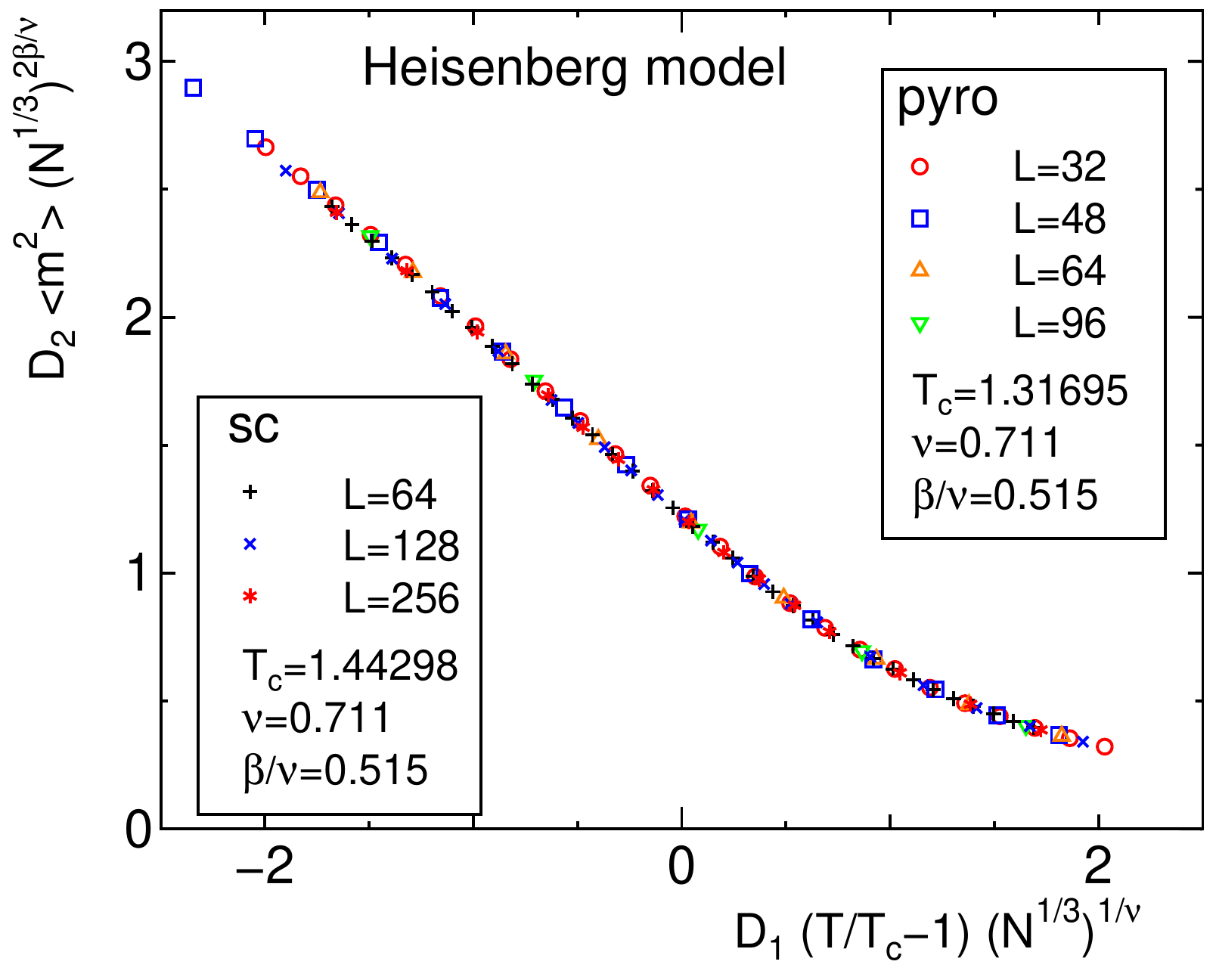}
\caption{\label{fig:fig16}
(Color online.) The universal FSS plot of the squared magnetization
of the Heisenberg model.
The system size $N$ is $16 L^3$ for the pyrochlore lattice, whereas it is
$L^3$ for the simple cubic lattice.
The choices of $T_c$, $\nu$, and $\beta/\nu$ are given in the figure.
The nonuniversal metric factors $D_1$ and $D_2$ are chosen as 1.0
for the simple cubic lattice.
The estimated $D_1$ and $D_2$ for the pyrochlore lattice are 0.92 and
0.91, respectively.
}
\end{center}
\end{figure}

We also give the universal FSS plots of the moment ratio and
of the squared magnetization of the Heisenberg model
in Figs.~\ref{fig:fig15} and \ref{fig:fig16}, respectively.
The nonuniversal metric factors
of the pyrochlore lattice are $D_1=0.92$ and $D_2=0.91$.

\section{Summary and Discussions}

We have performed high-performance computation of the ferromagnetic
classical spin models on the pyrochlore lattice.
We have determined the critical temperature accurately based on
the FSS of the Binder ratio.
The estimated critical temperatures are \IpyroTc, \XYpyroTc,
and \HpyroTc\ for the Ising model ($n=1$), the XY model ($n=2$),
and the Heisenberg model ($n=3$), respectively.
They are 93.4\%, 92.1\%, and 91.3\% of $T_c$ of the simple cubic
lattice, respectively.
Estimated critical exponents are universal ones of the 3D values.

Comparing with the data on the simple cubic lattice, we have argued
the universal FSS.  We have obtained nice universal FSS plots
by introducing nonuniversal metric factors $D_1$ and $D_2$.
If we choose $D_1$ and $D_2$ as 1.0 for the simple cubic lattice,
the nonuniversal metric factor $D_1$ for the pyrochlore
is 1.0, 0.96, and 0.92 for the Ising model, the XY model,
and the Heisenberg model, respectively;
the nonuniversal metric factor $D_2$ is 0.91 for all the spin models.

In the present work, we have treated the perfect lattice. 
The extension to random systems, for example, a diluted system, 
is straightforward. It will be left to a future study. 

We finally mention the algorithm of the present calculation.
We have used the single-GPU-based SW multi-cluster spin flip algorithm
\cite{komura15}
for large-scale simulations of spin models on the pyrochlore lattice.
This algorithm can be adapted to a variety of lattices
in a uniform fashion so long as a nearest-neighbor table is supplied.
We have again confirmed the efficiency of the algorithm.

\section*{Acknowledgments}

We thank Vitalii Kapitan and Yuriy Shevchenko for valuable discussions.
The results were obtained with using the equipment of Shared Resource Center
"Far Eastern Computing Resource" IACP FEB RAS and the
computer cluster of Far Eastern Federal University.
This work was supported by a Grant-in-Aid for Scientific Research
from the Japan Society for the Promotion of Science,
Grant Numbers JP25400406, JP16K05480.

\end{document}